\documentstyle[aps,prb,floats,epsfig]{revtex}

\begin{document}


\twocolumn[
\begin{center}
  \large\bf Atomic-scale simulations of nanocrystalline metals
\end{center}
\begin{center}
  J. Schi{\o}tz$^{1,2}$, T. Vegge$^{1,2}$, F. D. Di Tolla$^{1,3}$ and
  K. W. Jacobsen$^1$\\
  \emph{$^1$ Center for Atomic-scale Materials Physics (CAMP) and
  Department of Physics, Technical University of Denmark, DK-2800
  Lyngby, Denmark.\\
  $^2$ {\em Also at:\/} Materials Research Department, Ris{\o}
  National Laboratory, DK-4000 Roskilde, Denmark.\\
  $^3$ {\em Present address:\/} INFM and SISSA, Via Beirut 2--4,
  I-34014 Grignano (TS), Italy.
  }\\
  (Submitted February 11, 1999)
\end{center}

\begin{quote}
\begin{quote}\small
   Nanocrystalline metals, i.e.\ metals in which the grain size is in
  the nanometer range, have a range of technologically interesting
  properties including increased hardness and yield strength.  We
  present atomic-scale simulations of the plastic behavior of
  nanocrystalline copper.  The simulations show that the main
  deformation mode is sliding in the grain boundaries through a large
  number of uncorrelated events, where a few atoms (or a few tens of
  atoms) slide with respect to each other.  Little dislocation
  activity is seen in the grain interiors.  The localization of the
  deformation to the grain boundaries leads to a hardening as the
  grain size is increased (reverse Hall-Petch effect), implying a
  maximum in hardness for a grain size above the ones studied here.
  We investigate the effects of varying temperature, strain rate and
  porosity, and discuss the relation to recent experiments.  At
  increasing temperatures the material becomes softer in both the
  plastic and elastic regime.  Porosity in the samples result in a
  softening of the material, this may be a significant effect in many
  experiments.
\end{quote}
\end{quote}\vspace{1cm}
]


\section{INTRODUCTION}
\label{sec:intro}

The modeling of the mechanical properties of everyday materials is a
very challenging problem.  The main difficulty is the vastly different
length and time scales at which the various processes occur during
deformation --- ranging from the {\AA}ngstr\"om and sub-picosecond scales
of the atomic processes, to beyond the millimeter and second scales of
the macroscopic deformation.  Naturally, very different modeling
techniques are required to model phenomena at so different scales.
Atomic-scale simulations (typically molecular dynamics) can handle
time scales of up to a few nanoseconds and system sizes of up to $10^8$
atoms \cite{Ab97,BuAbKuDeYi98}, although one is typically limited to
significantly smaller system sizes and simulation times by the
available computer resources, and by the need to repeat the
simulations at different conditions.

At larger length and time scales, one is forced to abandon the
atomistic description of the material.  One option is dislocation
dynamics, where the fundamental unit of the simulation is the
dislocation.\cite{KuCaCoDePoBr92,DeKu97}  The fundamental idea of
dislocation dynamics is similar to that of molecular dynamics: by
calculating the forces on the dislocation (from the interactions with
each other as well as from any applied stress), the equations of
motions can be solved numerically.  The modeling becomes
significantly more complicated than molecular dynamics, mainly because
dislocations are lines whereas atoms are points.  For that reason,
dislocation dynamics is often done in two dimensions, where the
dislocations become point defects.\cite{BaCa95,ClGiNe97}  One
difficulty encountered when the atomic-scale description of the
material is abandoned is how to treat inherently atomic-scale
processes.  In the case of dislocation dynamics, it will mainly be
processes such as dislocation annihilation and core effects when the
dislocations are intersecting.  Parameters describing such processes
must be extracted from experiments, or calculated directly using
selected atomic-scale simulations.\cite{BuAbKuDeYi98,%
  ZhPrLoBe98,RaJaLePe97,RaJaLePeSrJo97}

At still larger length and time scales, continuum models must be
used.  This will typically be some kind of finite-element based
calculation, where the plastic behavior is described by a constitutive
relation.

A number of advanced simulation techniques have been proposed, where
simulation techniques on several length scales are combined into a
single multiscale simulation.  Various ways of combining finite
element calculations of the long-range elastic fields with
atomic-scale simulations of some regions of the material have appeared
in the
literature.\cite{KoGuFi91,Gu95,ThZhCaTe92,CaCaTh95,RaHeSiPaWo98,TaOrPh96}
Unfortunately, most of these methods are limited to quasi-static
simulations, i.e.\ to zero temperature.

Various simulation techniques have been proposed to extend the time
scales beyond the nanosecond scale reachable by molecular dynamics.
Some methods are based on ``accelerating'' the microscopic dynamics of
the system by modifying the potential energy surface\cite{Vo97} or by
extrapolating from higher to lower temperatures.\cite{VoSo99x} Other
methods, such as the ``nudged elastic band''
method\cite{MiJoSc95,JoMiJa98} and constrained molecular
dynamics,\cite{MuKrSwSc96,OtBr98,SpCi98} allow one to determine energy
barriers, from which reaction rates of slow
processes can be determined.

A different way of addressing the length-scale challenge is by
simulating suitably chosen problems, where the characteristic length
scale is within the reach of atomic-scale modeling.  This has the
advantage that the entire problem can be addressed at the atomic
scale, removing the need for the {\em a priori\/} assumptions about
which processes should be included in the simulation --- sometimes
unexpected atomic-scale processes are seen in such simulations,
processes that could not have been included in a coarser-scaled
simulation as its importance was not suspected.

One such atomic-scale problem is the mechanical deformation of
nanocrystalline metals, i.e.\ metals where the grain size is in the
nanometer range.  Nanocrystalline metals have recently received much
interest because they may have mechanical, chemical and physical
properties different from their coarse-grained counterparts.  For
example, the hardness and yield stress may increase 5--10 times when
the grain size is reduced from the macroscopic to the nanometer
range.\cite{SiFo94,Si97,MoMo97} 

Recently, computer simulations of the
structure\cite{Ch95,PhWoGl95,PhWoGl95b,ZhAv96,KePhWoGl97} of
nanocrystalline metals and semiconductors, and of their
elastic\cite{PhWoGl95} and plastic\cite{SwCa97,SwCa97b} properties
have appeared in the literature.  In previous papers, we described the
plastic deformation of nanocrystalline copper at zero
temperature.\cite{ScDiJa98,ScVeDiJa98} In this paper we focus on the
elastic and plastic properties of nanocrystalline metals, in particular
copper, at finite temperature.  We find that the materials have a very
high yield stress, and that the yield stress decrease with decreasing
grain size (reverse Hall-Petch effect).  The main deformation mode is
found to be localized 
sliding in the grain boundaries.

The high yield stress and hardness of nanocrystalline metals is
generally attributed to the Hall-Petch effect,\cite{Ha51,Pe53} where
the hardness increases with the inverse square root of the grain size.
The Hall-Petch effect is generally assumed to be caused by the grain
boundaries acting as barriers to the dislocation motion, thus
hardening the material.  The detailed mechanism behind this behavior
is still under debate.\cite{LiCh70}  A cessation or reversal of the
Hall-Petch effect will therefore limit the hardness and strength that
can be obtained in nanocrystalline metals by further refining of the
grain size.  There are a number of observations of a {\em reverse\/}
Hall-Petch effect, i.e.\ of a softening when the grain size is
reduced.\cite{ChRoKaGl89,FoWeSiKi92,MaKoMiMu98,YaScEhOg98}  The
interpretation of these results have generated some controversy.  It
is at present not clear if the experimentally reported reverse
Hall-Petch effect is an intrinsic effect or if it is caused by reduced
sample quality at the finest grain sizes.  The computer simulations
presented here show that an intrinsic effect is clearly possible.

The structure of the paper is as follows.  In section \ref{sec:sample}
we discuss the setup of the nanocrystalline model systems.  Section
\ref{sec:methods} discusses the simulation and analysis methods
used.  The simulation results are presented in section
\ref{sec:results}, and subsequently discussed in section \ref{sec:discuss}.

\section{SIMULATION SETUP}
\label{sec:sample}

In order to obtain realistic results in our simulations, and to be able
to compare our simulations with the available experimental data, we
have attempted to produce systems with realistic grain structures.
Unfortunately, the microscopic structure is not very well characterized
experimentally, and depends on the way the nanocrystalline metal was
prepared.  We have tried to create systems that mimic what is known
about the grain structure of nanocrystalline metals generated by inert
gas condensation.  The grains seem to be essentially equiaxed,
separated by narrow grain boundaries.  The grains are essentially
dislocation free.  The grain size distribution is
log-normal.\cite{SiFo94,Si94b,NaScHe98}

\subsection{Construction of the initial configuration}

In our simulations the grains are produced using a Voronoi
construction:\cite{OkBoSu92} a set of grain centers are chosen at
random, and the part of space closer to a given center than to any
other center is filled with atoms in a randomly rotated face-centered
cubic (fcc) lattice.  Periodic boundary conditions are imposed on the
computational cell.  This procedure generates systems without texture
and with random grain boundaries.  Effects of texture could
easily be included by introducing preferred orientations of the
grains.  In the limit of a large number of grains, the Voronoi
construction will generate a grain size distribution close to a
log-normal distribution.\cite{KuKuBaSh92}

In the grain boundaries thus generated, it is possible that two atoms
from two different grains get too close to each other.  In such cases
one of the atoms is removed to prevent unphysically large energies and
forces as the simulation is started.  To obtain more relaxed grain
boundaries the system is annealed for 10\,000 timesteps (50\,ps) at
300\,K, followed by an energy minimization.  This procedure is
important to allow unfavorable local atomic configurations to relax.

To investigate whether the parameters of the annealing procedure are
critical, we have annealed the same system for 50 and 100\,ps at
300\,K, and for 50\,ps at 600\,K.  We have compared the mechanical
properties of these systems with those of an identical system without
annealing.  We find that the annealing is important (the unannealed
system was softer), but the parameters of the annealing are
not important within the parameter space investigated.

A similar generation procedure has been reported by Chen,\cite{Ch95}
by D'Agostino and Van Swygenhoven ,\cite{AgSw96} and by Van
Swygenhoven and Caro.\cite{SwCa97,SwCa97b} A different approach was
proposed by Phillpot, Wolf and Gleiter:\cite{PhWoGl95,PhWoGl95b} a
nanocrystalline metal is generated by a computer simulation where a
liquid is solidified in the presence of crystal nuclei, i.e.\ small
spheres of atoms held fixed in crystalline positions.  The system was
then quenched, and the liquid crystallized around the seeds, thus
creating a nanocrystalline metal.  In the reported simulations, the
positions and orientations of the seeds were deterministically chosen
to produce eight grains of equal size and with known grain boundaries,
but the method can naturally be modified to allow randomly placed and
oriented seeds.  The main drawback of this procedure is the large
number of defects (mainly stacking faults) introduced in the grains by
the rapid solidification.  The stacking faults are clearly seen in the
resulting nanocrystalline metal (Fig.~7 of
Ref.~\onlinecite{PhWoGl95}).  The appearance of a large number of
stacking faults was also seen in the solidification of large clusters
even if the cooling is done as slowly as possible in atomistic
simulations.\cite{NiSeStJaNo94,Sc95}

\subsection{Structures}

A typical system (after annealing and cooling to zero Kelvin) with a
grain size of 5.2\,nm is shown in Fig.~\ref{fig:initial}.  The atoms
have been color coded according to the local crystal structure, as
determined by the Common Neighbor Analysis (see section
\ref{sec:analysis}).  In Figure \ref{fig:initialrdbf} the radial
distribution function (RDBF) $g(r)$ for the same system is shown.  It
is defined as the average number of atoms per volume at the distance
$r$ from a given atom.  The RDBF is seen to differ from that of a
perfect fcc crystal in two ways.  First, the peaks are not sharp
delta functions, but are broadened somewhat.  This broadening is in
part due to strain fields in the grains (probably originating from the
grain boundaries), and in part due to atoms in or near the grain
boundaries sitting close to (but not at) the lattice positions.  The
second difference is seen in the inset: the RDBF does not go to zero
between the peaks.  It is the signature of some disorder which in this
case comes from the grain boundaries.

Experimentally, information about the RDBF can be obtained from X-ray
absorption fine structure (XAFS) measurements.  This method has been used to
measure the average coordination number of Cu atoms in nanocrystalline
Cu, finding coordination numbers of $11.8 \pm 0.3$ and $11.9 \pm 0.3$
in samples with 34\,nm and 13\,nm grain size,
respectively.\cite{StSiNeSaHa95} From these results the average coordination
number of the atoms in the grain boundaries was estimated to $11.4 \pm
1.2$, i.e.\ within the experimental uncertainty it is the same as in
the bulk.  Integrating the first peak of the calculated RDBF
(Fig.~\ref{fig:initialrdbf}) gives an average
coordination number of $11.9\pm0.15$.  As the RDBF does not go to zero
between the first two peaks, it is not clear where the upper limit of
the integration should be chosen, hence the uncertainty.  The value
given is for an upper limit of 3.125\,\AA.  There is thus excellent
(but perhaps rather trivial)
agreement between the calculated and the experimental coordination
numbers.

Numerical studies have shown that the Voronoi construction results in
a grain size distribution that it is well described by a log-normal
distribution (although for more than 5000 grains a two-parameter gamma
distribution gives a better fit).\cite{KuKuBaSh92} In
Fig.~\ref{fig:lognormal} we show the grain size distributions in our
simulations with intended average grain sizes of 3.28 and 5.21
nanometers.  The observed distributions are consistent with log-normal
distributions, although due to the rather low number of grains it is
not possible to distinguish between a log-normal or a normal
distribution.

\section{SIMULATION METHODS}
\label{sec:methods}

We model the interactions between the atoms using a many-body
Effective Medium Theory (EMT) potential.\cite{JaNoPu87,JaStNo96} EMT
gives a realistic description of the metallic bonding, in particular
in fcc metals and their alloys.  Computationally, it is not much more
demanding than pair potentials, but due to its many-body nature it
gives a far more realistic description of the materials properties.

The systems can be deformed by rescaling the coordinates along a
direction in space (in the following referred to as the $z$
direction).  During this deformation either a conventional molecular
dynamics (MD) algorithm or a minimization algorithm is used to update
the atomic positions in response to the deformation.

In molecular dynamics, the timestep used when integrating the equation
of motion must be short compared to the typical phonon frequencies in
the system.  We use a timestep of 5\,fs, safely below the value where
the dynamics becomes unstable.  A consequence of the short timesteps
involved in MD simulations is that only brief periods of time can be
simulated.  For the size of systems discussed here 1\,ns (200000
timesteps) is for all practical purposes an upper limit, although for
repeated simulations 0.1\,ns is a more realistic limit.

One consequence of the short time-scale is that very high strain rates
are required to get any reasonable deformation within the available
time, a typical strain rate in the simulations reported here is $5 \times
10^8 s^{-1}$.  This is very high, but still the ends of the system
separates at velocities far below the speed of sound.  We have
investigated the effects of varying the strain rate, see section
\ref{sec:strainrate} .

As another consequence of the short time-scale, slower processes will
not be seen in the simulations.  In particular, most diffusional
processes will be unobservable (this would also be the case in
experiments performed at these high strain rates).  However,
measurements of diffusional creep (Coble creep) in nanocrystalline
metals indicate that diffusional creep is not a large
effect.\cite{NiWeSi91,NiWa91}

Some of the simulations were performed using a minimization procedure,
i.e.\ the system is kept in a local energy minimum while it is
deformed.  In such a simulation time is not defined, since we are not
solving an equation of motion.  In a similar way, time should not be
relevant in an experiment performed truly at zero temperature, since
there will be no thermally activated processes, and thus no way the
sample can leave a local energy minimum (neglecting quantum
tunelling).  So the strain rate will not have an effect on the
results, providing it is low enough to prevent a heating of the
sample, and providing the minimization procedure is fully converged.
The minimization simulations can thus be seen as a model for idealized
experiments at zero temperature in the low strain rate limit, where
there is time for the heat generated by the deformation to be removed.

\subsection{The minimization procedure}
\label{sec:zerotemp}

To simulate deformation at zero temperature a minimization procedure
was used to keep the system in or near a local minimum in energy at
all times.  The deformation and minimization was done simultaneously.
The minimization algorithm is a modified molecular dynamics
simulation.\cite{footnote1}
After each MD
timestep the dot product between the momentum and the force is
calculated for each atom.  Any atom where the dot product is negative
gets its momentum zeroed, as it is moving in a direction where the
potential energy is increasing.  In this way the kinetic energy of an
atom is removed when the potential energy is close to a local minimum
along the direction the atom is moving.  This minimization procedure
quickly brings the system close to a local minimum in energy, but a
full convergence is not obtained, as it would require a number of
timesteps at least as great as the number of degrees of freedom in the
system.  However, we find only little change in the development of the
system, when we increase the number of minimization steps.

At each timestep the system is deformed by a tiny scaling of the
coordinates, the $z$ coordinates are multiplied by $1 + \epsilon$, the
$x$ and $y$ coordinates by $1 - \nu \epsilon$, where $\epsilon$ is a
very small number, chosen to produce the desired deformation
rate.\cite{footnote2}
The constant $\nu$ is an ``approximate
Poisson's ratio''.  A Monte Carlo algorithm is used to optimize the two
lateral dimensions of the system: after every 20th timestep a change
in the lateral dimensions is proposed.  If the change result in a reduction of
the energy it is accepted, otherwise it is discarded.  In this way the
exact value chosen for $\nu$ becomes uncritical, as the Monte Carlo
algorithm governs the contraction in the lateral directions.  The use
of $\nu \neq 0$ is just for computational efficiency.  We used $\nu =
0.4$ as this was between the optimal value in the part of the
simulation where elastic deformation dominates (Poisson's ratio
$\approx 0.3$) and in the part where plastic deformation dominates ($ \nu
\approx 0.5$ as volume is then conserved).

A few simulations were repeated using a Conjugate Gradient (CG)
minimization instead of the MD minimization algorithm.  The two
algorithms were approximately equally efficient in these simulations,
provided that the CG algorithm was restarted approximately every 20
line minimizations.  Otherwise the CG algorithm will not minimize
twice along the same direction in the $3N$-dimensional configuration
space.  

\subsection{Molecular dynamics at finite temperature}
\label{sec:hitemp}

At finite temperatures a conventional molecular dynamics algorithm is
used, where the Newtonian equations of motion for the atoms are solved
numerically.  During the simulation the
computational box was stretched as described above.  The Monte Carlo
algorithm optimizing the lateral dimensions is the conventional
Metropolis algorithm.

Before the deformation is applied the system is heated to the desired
temperature by a short Molecular Dynamics simulations using Langevin
dynamics,\cite{AlTi87} i.e.\ where a friction and a fluctuating force is added
to the equation of motion of the atoms.  When the desired temperature
has been reached (after approximately 10 ps), the simulations is
performed using the velocity Verlet algorithm.\cite{SwAnBeWi82}  During the
deformation process the internal energy increases by the work
performed on the system.  This amounts in practice to a small heating
of the system of the order of $\sim$30 K.

\subsection{Analysis of the results}
\label{sec:analysis}

While the simulation is performed, the stress field is regularly
computed.  The stress tensor is the derivative of the
free energy of the system with respect to the strain.  The
effective medium theory allows us to define an energy per atom, which
allows us to define an ``atomic'' stress for each atom.  The stress is
a suitable derivative of the energy with respect to the interatomic
distances:\cite{EgMaVi80,RaRa84,RaRa85}
\begin{equation}
  \label{eq:stress}
  \sigma_{i,\alpha\beta} = \frac{1}{v_i} \left( -\frac{p_{i,\alpha}
    p_{i,\beta}}{m_i} + \frac{1}{2} \sum_{j \ne i} \frac{\partial
    E_{\text{pot}}}{\partial r_{ij}} \frac{r_{ij,\alpha}
    r_{ij,\beta}}{r_{ij}} \right)
\end{equation}
where $\sigma_{i,\alpha\beta}$ is the $\alpha,\beta$ component of the
stress tensor for atom $i$, $v_i$ is the volume assigned to atom $i$
($\sum_i v_i = V$, where $V$ is the total volume of the system), $m_i$
is the mass of atom $i$, $p_{i,\alpha}$ is the $\alpha$ component of
its momentum and $r_{ij}$ is the distance between atoms $i$ and $j$
($r_{ij,\alpha}$ is a component of the vector from atom $i$ to $j$).

The atomic stress tensor cannot be uniquely defined.
Eq.~\ref{eq:stress} is based on the Virial Theorem, but other
definitions are possible.\cite{Ts79,Lu88,ChYi91} When the atomic stress is
averaged over a region of space the various definitions quickly
converge to a macroscopic stress field.\cite{Lu88,ChYi91}

During the simulation, stress-strain curves are calculated by averaging
the atomic stresses over the entire system.

To facilitate the analysis of the simulations the local atomic order
was examined using an algorithm known as Common Neighbor Analysis
(CNA).\cite{JoAn88,ClJo93}  In this algorithm the bonds between an
atom and its neighbors are examined to determine the crystal structure
(two atoms are considered to be bonded if they are closer together
than a cutoff distance chosen between the first two peaks in the
radial distribution function).  Bonds are classified by three
integers $(ijk)$.  The first integer $i$ is the number of {\em common
  neighbors\/}, i.e. atoms bonded to both atoms in the bond under
consideration.  The second integer $j$ is the number of bonds between
these common neighbors.  The third integer $k$ is the longest chain
that can be formed by these bonds.

The number and type $(ijk)$ of bonds that an atom has determines the
local crystal structure.  For example, atoms in a perfect fcc crystal
have 12 bonds of type 421, whereas atoms in a perfect hcp crystal have
six bonds of type 421 and six of type 422.  In these simulations we
have mainly used CNA to classify atoms into three classes: fcc, hcp
and ``other'', where the ``other'' class is atoms that have a
different number of bonds than 12, or that have bonds that are not of
type 421 or 422.

The use of CNA makes dislocations, grain boundaries and stacking
faults visible in the simulation.  Intrinsic stacking faults appear as
two adjacent ${111}$ planes of hcp atoms, extrinsic stacking faults
are two ${111}$ planes of hcp atoms separated by a single ${111}$
plane of fcc atoms, whereas twin boundaries will be seen as a single
${111}$ plane of hcp atoms.  Dislocation cores and grain boundaries
consist of atoms in the ``other'' class, although grain boundaries
also contain a low number of hcp atoms.

When analyzing simulations made at finite temperatures, the lattice
vibrations may interfere with the CNA.  For simulations at
temperatures up to 300\,K it is sufficient to choose the cutoff
distance carefully, but for higher temperatures it is necessary to
precede the CNA analysis by a short minimization (short enough to
remove most of the lattice vibrations, but not long enough to allow
dislocations etc.\ to move).  For consistency, such a minimization
procedure was used for all finite-temperature simulations regardless
of temperature.

\section{RESULTS}
\label{sec:results}

In this section we first discuss the results of the simulations based
on the minimization algorithm (i.e.\ the zero-temperature results), and
then we discuss the results obtained using molecular dynamics at
finite temperatures.

\subsection{Simulations at zero temperature}

During the deformation, we calculate the average stress in the system
as a function of the strain.  Fig.~\ref{fig:stressstrain0K} shows
the obtained stress-strain curves from simulations at 0\,K.  We see a
linear elastic region followed by a plastic region with almost
constant stress.  Similar results are found for palladium
(Fig.~\ref{fig:stressstrainPd}).  Each stress-strain curve shown in
Figs.~\ref{fig:stressstrain0K} and \ref{fig:stressstrainPd} are
obtained by averaging over a number of simulations with different
(randomly produced) grain structures with the same average grain
size.  A set of stress-strain curves from individual simulations are
shown in Fig.~\ref{fig:stressstrainIndiv}.

One of the rationales using a minimization procedure to study the
deformations was the hope that the system would evolve through a
series of local energy minima, separated by discrete events when the
applied deformation causes the minima to disappear.  In this way, the
simulation would have resulted in a unique deformation history for any
given sample.  However, the deformation turned out to happen through a
very large number of very small processes, that could not be
individually resolved by this procedure (see below).  One symptom of
this is, that the individual curves in
Fig.~\ref{fig:stressstrainIndiv} are not completely reproducible.  Any
even minor change in the minimization procedure, or a perturbation of
the atomic coordinates, will result in a slightly different path
through configuration space, and in different fluctuations in the
stress-strain curves.  Those differences are suppressed when average
stress-strain curves are calculated, as in
Fig.~\ref{fig:stressstrain0K}, and would also disappear as the system
size (and thus the number of grains) increase.

\subsubsection{Young's Modulus}

In the linear elastic region the Young's modulus is found to be around
90-105\,GPa and it is increasing with increasing grain size.  The
experimental value for macrocrystalline copper is 124\,GPa at
300\,K,\cite{GS64} and the value found for single crystals using this
potential is 150\,GPa at 0\,K (Hill average calculated from the
anisotropic elastic constants $C_{11} = 173$\,GPa, $C_{12} =
116$\,GPa, $C_{44} = 91$\,GPa).  A similar reduction of Young's
modulus is seen in simulations of nanocrystalline metals grown from a
molten phase.\cite{PhWoGl95} The low value is due to the large volume
fraction of the atoms being in the grain boundaries.\cite{ShKoTsPh95}
These atoms experience a different atomic environment, which could
result in a reduction of the elastic moduli similar to what is seen in
amorphous metals.  This local reduction of the elastic constants in
grain boundaries is confirmed by atomistic
simulations.\cite{KlWoLuPh90}

Experimental measurements of the Young's modulus of high-quality
(i.e.\ low-porosity) samples of nanocrystalline copper and palladium
show a reduction in Young's modulus of at most a few percent when
correcting for the remaining porosity.\cite{SaYoWe97} These results
were obtained for significantly larger grain sizes than were used in
the simulations.  The reduction of Young's modulus that we observe in
these simulation, will be difficult to detect experimentally due to
the much lower volume fraction of atoms in the grain boundaries for
typical grain sizes in high-quality samples ($\gtrsim 20$ nm).

\subsubsection{Yield and flow stress}

The onset of plastic deformation is usually described by the yield
stress $\sigma_y$, traditionally defined as the stress where the
strain is 0.002 larger than what would be expected from extrapolation
from the elastic region.  In these simulations the stress continues to
increase after the yield point has been reached, until it reaches a
plateau and becomes constant (or slightly decreasing).  We call the
level of the plateau the flow stress.

Fig.~\ref{fig:hallpetch0K} shows the dependence of the yield and flow
stress on the grain size.  A clear reverse Hall-Petch effect is
observed, i.e. a softening of the material as the grain size is
reduced, as  discussed in a previous paper.\cite{ScDiJa98}

\subsubsection{Structural changes}

Fig.~\ref{fig:last} shows the same system as Fig.~\ref{fig:initial},
but after 10\,\% deformation.  Some stacking faults have appeared in
the grains, they are caused by partial dislocations (Shockley
partials) nucleating at the grain boundaries and moving through the
grains.  One such dislocation is seen in the figure.

The radial distribution function (Fig.~\ref{fig:lastrdbf}) has been
changed somewhat by the deformation.  The peaks have been broadened,
this is mainly caused by the anisotropic stress fields in the sample.
The ``background level'' between the first two peaks has increased a
little,   indicating a larger amount of disorder in the system.
Increased disorder is also seen in Fig.~\ref{fig:last}, where the
grain boundaries appear to have become slightly thicker compared to
the initial configuration. This is confirmed by
Fig.~\ref{fig:countatoms}, showing the number of atoms in different
local configurations before and after the deformation.  We see how
more atoms are neither fcc nor hcp after the deformation than before.

A strong increase in the number of atoms near stacking faults (atoms
in hcp symmetry) is also seen in Figs.~\ref{fig:last} and
\ref{fig:countatoms}.  The stacking faults appear as partial
dislocations move through the system, and they are thus the signature
of dislocation activity.  At zero temperature, we do not observe cases
where a second partial dislocation erases the stacking faults (we
observe only a very few atoms changing from local hcp order to local
fcc order).  We can therefore use the total number of hcp-ordered
atoms to estimate an upper bound on the amount of plastic deformation
caused by the dislocations.

If a dislocation with Burgers vector $\vec{b}$ runs through the entire
system, the dimensions of the system are changed by $\vec{b}$ and the
strain $\varepsilon_{zz}$ is thus $b_z L_z^{-1}$, where $b_z$ is the $z$
component of the Burgers vector and $L_z$ is the dimension of the
system in the $z$ direction.  If the dislocation only passes through a
part of the system, the resulting deformation is reduced by $A \cos
\phi (L_x L_y)^{-1}$, where $A$ is the area of the slip, $L_x$ and
$L_y$ are the lateral dimensions of the system and $\phi$ is the angle
between the slip plane and the $xy$ plane.
The contribution from a slip plane to the $zz$ component of the strain
is thus $\varepsilon_{zz} = (b_z L_z^{-1}) A \cos \phi (L_x L_y)^{-1}$.

The maximal value of $b_z$ is $b \sin \phi$, where $b = | \vec{b} |$,
since $\vec{b}$ lies in the slip plane.  The maximal strain from the
slip thus becomes $\varepsilon_{\text{max}} = b A (2 V)^{-1} = a A (2
\sqrt{6} V)^{-1}$ for $\phi = \pi/4$, as the Burgers vector of a
Shockley partial is $b = a / \sqrt{6}$, where $a$ is the lattice
constant and $V$ is the volume of the simulation cell.  A slip plane
of area $A$ results in two $\{111\}$ planes of hcp atoms, i.e.\ %
$4 A (\sqrt{3} a^2)^{-1}$ atoms.  The total system contains $4 V
a^{-3}$ atoms, so the fraction of hcp atoms becomes $n = A a (\sqrt{6}
V)^{-1}$.  Hence a fraction $n$ of hcp atoms can at most have resulted
in a strain of $\varepsilon_{\text{max}} = 2^{-3/2} n$.  As the
simulation generate at most 9\% hcp atoms during 10\% deformation, we
get that $\varepsilon_{\text{max}} \approx 3\%$, provided that all
slip planes and Burgers vectors are ideally aligned.  It is therefore
clear that the main deformation mode is not by dislocation motion.

Fig.~\ref{fig:motion} illustrates how the main part of deformation has
taken place.  The atoms are colored according to their motion {\em
  relative to the global stretching of the system}.  We clearly see
that the upper parts of the grains have moved down and the lower parts
up, relative to what would be expected in a homogeneous deformation.
This shows that the grains do not stretch as much as in a homogeneous
deformation.  On the other hand, it is seen that significant
deformation has happened in the grain boundaries, as the atoms
typically are moving up on one side and down on the other side of a
grain boundary.  An analysis of that deformation shows that it is in
the form of a large number of apparently uncorrelated small slipping
events, where a few atoms (or a few tens of atoms) move relatively to
reach other,\cite{ScDiJa98} i.e.\ not in the form of collective motion
in the grain boundaries.  A minor part of the plastic deformation is
in the form of dislocation motion inside the grains.  The slip planes
are clearly seen in Fig.~\ref{fig:motion}, in particular in the large
grain in the upper left part of the figure, where two dislocations
have moved through the grain, and a third is on its way near one of
the previous slip planes.

\subsection{Simulations at finite temperature}

The simulations were repeated with the same systems (i.e. the same
initial grain structure) at finite
temperatures.  Fig.~\ref{fig:sstemp} shows the stress-strain curves
for the same system at different temperatures.  We clearly see a
softening with increasing temperature.  Rather large fluctuations are
seen in the curves.  These are mainly thermal fluctuations and
fluctuations due to single major ``events'' in the systems (e.g. the
nucleation of a dislocation).  These fluctuations are only visible due
to the small system size.

The softening of the material with decreasing grain size is also
observed in simulations at 300\,K, see
Fig.~\ref{fig:stressstrain300K}. 

\subsubsection{Young's Modulus}
\label{sec:youngfinite}

Young's Modulus ($E$) is the slope of the stress-strain curve in the
linear region.  When calculating Young's modulus from the simulation
data, a compromise must be made between getting enough data point for
a reliable fit, and staying within the clearly linear region.  For the
zero-temperature simulations, fitting Young's modulus to the data
point for $\varepsilon < 0.1\%$ satisfies both conditions, but for the
finite-temperature simulations more data points are required.  We have
chosen to use data in the interval $\varepsilon < 1\%$, this ensures
that we have enough data for a reliable fit, but results in a slight
underestimate of the Young's modulus, as some plastic deformation is
beginning in this interval.  The results of this procedure is shown in
Fig.~\ref{fig:youngtemp}, showing Young's modulus for a single system
with grain size $d = 5.2$\,nm,
simulated at different temperatures.  For consistency, the larger
strain interval has been used even for the $T = 0$ simulation.  Using
the smaller interval ($\varepsilon < 0.1\%$) would result in $E =
119$\,Gpa instead of $E = 100$\,GPa.

The observed temperature dependence of $E$ is approximately -72 MPa/K,
which is somewhat larger than what has been observed experimentally
(-40 MPa/K) in copper with a grain size of 200\,nm.\cite{LeBuRoKoFiGr95}
This may be
because the Young's modulus of the grain boundaries is more
temperature sensitive than in the bulk, or it may be due to increased
creep in the higher-temperature simulations, see the discussion in
section \ref{sec:strainrate}.

\subsubsection{Yield and flow stress}

The Yield stress is again determined as the stress where the strain is
0.2\% above what would be expected from extrapolation from the linear
regime.  The difficulties leading to an underestimate of the Young's
Modulus thus leads to an overestimate of the yield stress.  The values
for the yield stress obtained for the simulations at 300\,K can
therefore not be compared directly with the values obtained at 0\,K,
but values obtained at different grain sizes can of course be
compared, as the same method was used to estimate the yield stress in
all cases.
The flow stress is a far more well-defined quantity, and direct
comparison is possible between simulations at different temperatures.
The variation of the yield and flow stresses with temperature is seen
in Fig.~\ref{fig:flowtemp}.

Fig.~\ref{fig:hallpetch300K} shows the variation of both the yield-
and the flow stress with grain size.  As in the simulations at 0\,K a
reverse Hall-Petch relationship is found.

\subsubsection{Structural changes}

The main deformation mode appears to be the same at zero and at finite
temperatures.  Figure \ref{fig:motion300K} shows the atomic displacements,
again the majority of the deformation has taken place in the grain
boundaries, only a few slip planes are seen.  

The grain boundaries do not appear to increase as much in thickness as
they to at 0\,K.  Fig.~\ref{fig:classdevel} shows the change in the
fraction of atoms in different local environments during the
deformation of the same system at 0\,K and at 300\,K.  In both cases
we clearly see an increase in the number of hcp atoms (stacking
faults) due to the motion of dislocations through the grains, but the
number of atoms in the grain boundaries increases significantly more
at 0\,K than at 300\,K (the increase is approximately twice as big at
0\,K as at 300\,K).  The increase appears to be caused by the
deformation in the grain boundaries.  Apparently, the local disorder
introduced in this way is partially annealed out at 300\,K.

\subsubsection{Strain rate}
\label{sec:strainrate}

The finite-temperature simulations presented in this paper were
performed at a strain rate of $\dot\varepsilon = 5\times10^8 s^{-1}$,
unless otherwise is mentioned.

In order to investigate the influence of the strain rate, we simulated
the same deformation of the same system using different strain rates
in the range $2.5\times10^7 s^{-1}$ -- $1.0 \times 10^{10} s^{-1}$,
see Fig.~\ref{fig:strainrate}.  A strong dependence on the strain rate
is seen for strain rates above $1\times10^9 s^{-1}$.  Below this
``critical strain rate'' the strain rate dependence on the
stress-strain curves is far less pronounced.
Fig.~\ref{fig:sratesummary} confirms this impression.  It shows the
yield and flow stress as a function of the strain rate.  Experiments
on ultrafine-grained ($d \approx 300$ nm) Cu and Ni show a clear
strain rate dependence on the yield and flow stresses at high strain
rates.\cite{GrLoCaVaAl97} 

Perhaps surprisingly, the Young's modulus appears to depend on the
strain rate as well (Fig.~\ref{fig:strainrate}).  This indicates that
some kind of plastic deformation occurs in the ``linear elastic''
region.  This is confirmed by stopping a simulation while
the system still appears to be in the elastic region, and then
allowing it to contract until the stresses are zero.  The system does
not regain the original length: plastic deformation has occurred.

To examine the time-scale over which this deformation
occurs, a configuration was extracted from the simulation at
$\dot\varepsilon = 2.5 \times 10^7 s^{-1}$ after 0.4\% deformation.
The system was held at a fixed length for 300 ps while the stress is
monitored, see Fig.~\ref{fig:relaxation}.  The stress is seen to
decrease with a characteristic time of approximately 100 ps.  By
plotting the atomic motion in a plot similar to Figs.~\ref{fig:motion}
and \ref{fig:motion300K}, it is seen that the relaxation is due to
small amounts of plastic deformation {\em in the grain boundaries\/}.
The consequence of this is that the systems do not have time to relax
completely during the simulations, explaining the observed strain rate
dependence.  In order to allow for complete relaxation of the systems,
strain rates far below what is practically possible with MD
simulations are required.

\subsubsection{Grain rotation}

Grain rotation has previously been reported in simulations of
nanocrystalline nickel.\cite{SwCa97b} We have investigated the
rotation of the grains during some of the simulations, the results are
summarized in Fig.~\ref{fig:rotation}.  The figure shows the rotation
of five randomly selected grains as a function of strain and
temperature.  The rotations were identified by a three-dimensional
Fourier transform of the positions of the atoms in the grains.

We see that the grain rotation increases with increasing temperature.
There is a large variation between how much the individual grains
rotate.  The grains with the largest rotations keep the same axis of
rotation during the entire deformation, whereas the grains that only
rotate a little have a varying axis of rotation.  Probably some grains are
in a local environment where a significant rotation results in an
advantageous deformation of the sample which reduces the stress.
Other grains are randomly rotated as the many small deformation
processes in the grain boundaries occur.

\subsubsection{Porosity}
\label{sec:porosity}

As the observed reverse Hall-Petch effect is often explained as an
artifact of sample porosity (see section \ref{sec:discHP}), we found
it relevant to study how pores influence the mechanical properties.
The void structure in experimentally produced samples is usually not well
known, so we chose to study several different types of voids.  In all
cases the voids resulted in a reduction of both the Young's modulus
and the flow stress, see Fig.~\ref{fig:voids}.

\paragraph*{Elliptical voids.}

These crack-like voids were created by removing all atoms within an
oblate ellipsoid with an axis ratio of 3.16.  The short axis can be
oriented along the pulling direction (the $z$-axis) or perpendicular
to it (the $x$-axis).  The former orientation corresponds to cracks
that are activated by the applied stress field, the effect of these
cracks is therefore expected to be much larger than the effect of the
``inactive'' cracks.  This is clearly seen in Fig.~\ref{fig:voids}.

\paragraph*{Missing grains.}

There have been reports of pore sizes comparable to (and proportional
to) the grain size.\cite{SaYoWe97,SaEaWe98}  To emulate this, we have
tried to remove whole grains from the system.  As the grains are
approximately equiaxed, it is not surprising that the effect of
removing a grain is intermediate between the effects of removing
ellipsoids in the two orientations, provided that approximately the
same number of atoms are removed.

\paragraph*{Missing grain boundary atoms.}

In samples experimentally produced by compacting a powder, it is
reasonable to assume that the porosity will mainly be in the form of
(possibly gas-filled) voids between the grains.  There is also some
experimental evidence that this is indeed the
case.\cite{SaEaWe98,AgElYoHeWe98}  To emulate this, we have removed
all atoms in the grain boundaries within one or nine spherical regions
in the sample, creating one large or nine small voids in the grain
boundaries.  This type of voids have the largest effect on the
materials properties, giving a reduction of 35--40\% in the Young's
modulus and flow stress for a 12.5\% porosity.
It seems rather natural that a large effect is obtained with the voids
concentrated in the grain boundaries since we know that the main part
of the deformation is carried by these boundaries.

\section{Discussion}
\label{sec:discuss}

\subsection{The reverse Hall-Petch effect}
\label{sec:discHP}

A reverse Hall-Petch effect in nanocrystalline copper was first
observed in nanocrystalline Cu and Pd by Chokshi \emph{et al.} in
1989.\cite{ChRoKaGl89}  Since then, there have been numerous
observations of softening at very small grain
sizes.\cite{FoWeSiKi92,MaKoMiMu98,YaScEhOg98} 

The reverse Hall-Petch effect seems to depend strongly on the sample
preparation technique used and on the sample history, perhaps indicating
that in most cases the reverse Hall-Petch effect is caused by various
kinds of defects in the samples.  Surface defects alone have been
shown to be able to decrease the strength of nanocrystalline metals by
a factor of five,\cite{NiWeSi91,We93} and recent studies have shown
that even very small amounts of porosity can have a dramatic effect on
the strength.\cite{SaYoWe97,SaEaWe97}  Improved inert gas condensation
techniques\cite{SaFoThEaWe97} have reduced the porosity resulting in
samples with densities above 98\% of the fully-dense value.  In these
samples the ordinary Hall-Petch effect is seen to continue down to
grain sizes around 15\,nm.\cite{SaYoWe97}  There are only few data
points below that grain size, but apparently no further increase in
the hardness is seen.  It is suggested that most of the
observations of a reverse Hall-Petch effect in nanocrystalline metals
are a result of poor sample quality.\cite{SaYoWe97}  This impression
is supported by literature studies\cite{FoWeSiKi92,Gr93} indicating
that the reverse Hall-Petch effect is mainly seen when the grain size
is varied by repeated annealing of a single sample, whereas an
ordinary Hall-Petch relationship is seen when as-prepared samples are
used.

However, there does seem to be a deviation from the Hall-Petch effect
for grain sizes below approximately 15\,nm, where the Hall-Petch slope
is seen to decrease or vanish in samples produced with various
techniques.  This is seen in Cu samples produced by inert gas
condensation followed by warm compaction (sample densities above
98\%)\cite{SaYoWe97} and in electroplated Ni (claimed to be porosity
free).\cite{ElErPaAu92}

There are theoretical arguments for expecting that the Hall-Petch
relation ceases to be valid for grain sizes below $\sim$20\,nm: as the
grain size becomes too small, dislocation pileups are no longer
possible, and the usual explanation for Hall-Petch behavior does not
apply.\cite{NiWa91,PaMaAr93,WaWaAuEr95}

Many models have been proposed to explain why a reverse Hall-Petch
effect is sometimes seen.  Chokshi \emph{et al.}\cite{ChRoKaGl89}
proposed that enhanced Coble creep, i.e.\ creep by diffusion in the
grain boundaries, should result in a softening at the smallest grain
sizes as the creep rate increases with decreasing grain size ($d$) as
$d^{-3}$.  Direct measurements of the creep rate have however ruled
this out.\cite{NiWa91,NiWeSi90}

It has been suggested that the grain boundaries in nanocrystalline
metals have a different structure, making them more transparent to
dislocations than ``ordinary'' grain
boundaries.\cite{VaChBoKaBa92,LiBaNa93,LuSu93}  If it becomes possible
for the dislocations to run through several grains as the grain size is
reduced, the Hall-Petch relations would break down.  In our
simulations, we have not observed dislocations moving through more
than one grain.

If the Hall-Petch effect is explained by appealing to dislocation
sources in the grain boundaries, the Hall-Petch relationship is
expected to break down when the grain sizes becomes so low that there
are no longer dislocation sources in all grain
boundaries\cite{LiSuWa94} (assuming a constant density of dislocation
sources in the grain boundaries).

Hahn \emph{et al.}\cite{HaPa97,HaMoPa97} suggest that the reverse Hall-Petch
effect is caused by deformation in the grain boundaries.  If a grain
boundary slides, stress concentrations build up where the grain
boundary ends, limiting further sliding.  Substantial sliding on a
macroscopic scale occurs when sliding occurs on slide planes
consisting of many aligned grain boundaries; the sliding is hindered
by the roughness of the slide plane due to its consisting of many
grain boundaries.   As the grain size is reduced and becomes
comparable to the grain boundary width, the roughness of
such slide planes decreases and the stress required for mesoscopic
sliding decreases.  This would result in a reverse Hall-Petch effect.
They estimate the transition from normal to reverse Hall-Petch effect
to occur at grain sizes near 50\,nm for Cu.\cite{HaPa97}  

The simulations reported in the present paper indicate that the main
deformation mechanism at these grain sizes is indeed sliding in the
grain boundaries.  However, it is not clear if the proposed
``collective'' sliding events are occurring, it appears that sliding
occurs on individual grain boundaries, and that the resulting stress
buildup is relieved through dislocation motion in the grains.  There
is a competition between the ordinary deformation mode (dislocations)
and the grain boundary sliding.  As the grain size is increased, the
dislocation motion is eventually expected to dominate, and we expect a
transition to a behavior more like what is seen in coarse-grained
materials, incl.\ a normal Hall-Petch effect.  The transition is
beyond what can currently be simulated at the atomic scale, but we do
see a weak increase in the dislocation activity when the grain size is
increased: The increase in the fraction of hcp atoms during a simulation
is increasing slightly with the grain size (Fig.~\ref{fig:countatoms}).

\section{CONCLUSIONS}
\label{sec:concl}

Molecular dynamics and related techniques have been shown to be a
useful approach to study the behavior of nanocrystalline metals.  We
have in detail investigated the plastic deformation of nanocrystalline
copper, and shown that the main deformation mode is sliding in the
grain boundaries.  The sliding happens through a large number of
small, apparently uncorrelated events, where a few grain boundary
atoms (or a few tens of atoms) move past each other.  It remains the main
deformation mechanism at all grain sizes studied (up to 13 nm), even
at zero temperature.  As the grain boundaries are the main carriers of
the deformation, decreasing the number of grain boundaries by
increasing the grain size leads to a hardening of the material, a
\emph{reverse Hall-Petch effect}.  This is observed in the
simulations, both for $T=0K$ and for $T=300K$.

The Young's moduli of the nanocrystalline systems are found to be
reduced somewhat compared to the experimental value for
polycrystalline copper with macroscopic grain sizes, decreasing with
decreasing grain size.  This indicates that the grain boundaries are
elastically softer than the grain interiors.  The Young's modulus is
decreasing with increasing temperature at a rate somewhat above what
is seen experimentally in coarser-grained copper.

Pores in the samples have a large effect on both the Young's modulus
and the flow stress.   This effect is enhanced if the pores are mainly
in the grain boundaries, as one could expect in samples produced
experimentally by inert gas condensation.  Sample porosity can explain
a large number of experiments showing reverse Hall-Petch effect, but
the softening due to grain boundary sliding may be important for
high-quality samples with grain sizes close to the lower limit of what
can be reached experimentally.

\section*{ACKNOWLEDGMENTS}

Major parts of this work was financed by The Danish Technical Research
Council (STVF) through Grant No.~9601119.  Parallel computer time was
financed by the Danish Research Councils through Grant No.~9501775.
The Center for Atomic-scale Materials Physics is sponsored by the
Danish National Research Council.


\begin{thebibliography}{10}
\bibitem{footnote2}
In order to implement periodic boundary conditions all atomic coordinates are
  internally represented in a ``scaled space'', where coordinates are between 0
  and 1. The real coordinates are obtained by multiplying with a matrix $M$.
  The system is deformed by changing the relevant components of $M$, not by
  changing the scaled-space coordinates.
\end{thebibliography}


\newpage
\newlength{\figwidth}
\setlength{\figwidth}{3.375in}

\begin{figure}
  \begin{center}
    \leavevmode
    \epsfig{file=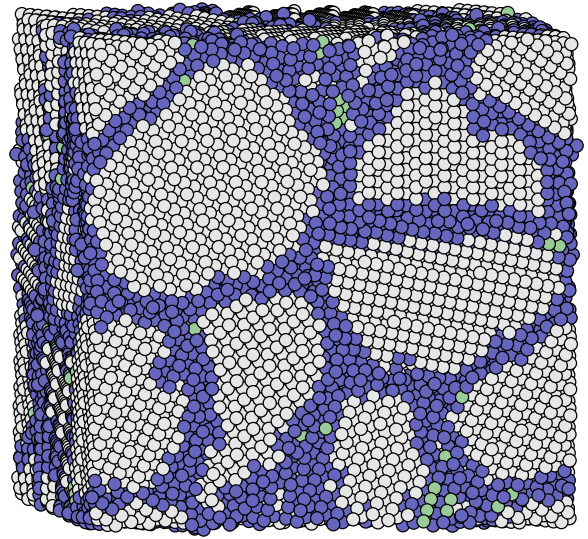, clip=, width=\figwidth}
  \end{center}
  \caption{The initial configuration of a nanocrystalline copper
    system.  The system contains approximately 100\,000 atoms arranged
    in 16 grains, giving an average grain diameter of 5.2\,nm.  Atoms
    are color-coded according to the local crystalline order.  White
    atoms are in a perfect fcc environment.  Green atoms are in local
    hcp order, which for example corresponds to stacking faults.
    Atoms in any other environment (in grain boundaries and 
    dislocation cores) are colored purple.}
  \label{fig:initial}
\end{figure}

\begin{figure}
  \begin{center}
    \leavevmode
    \epsfig{file=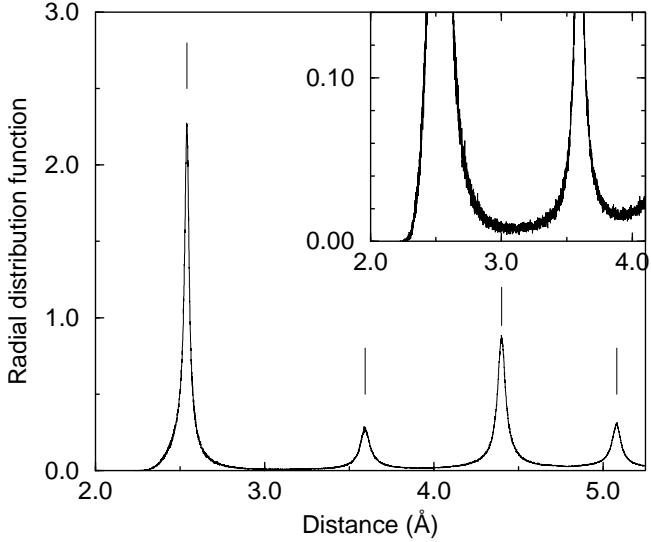, angle=-90, width=\figwidth, clip=}
  \end{center}
  \caption{The radial distribution function $g(r)$ for the system in
    Fig.~\ref{fig:initial} at $T=0$.  It is the average number of
    atoms per volume at the distance $r$ from a given atom.  The marks
    above the peaks indicate the positions of the (infinitely sharp)
    peaks in a perfect crystal at $T=0$.  The inset shows that $g(r)$
    does not vanish between the first and second peak as it does for a
    perfect crystal.  This contribution comes from the grain
    boundaries.}
  \label{fig:initialrdbf}
\end{figure}

\begin{figure}[htbp]
  \begin{center}
    \leavevmode
    \epsfig{file=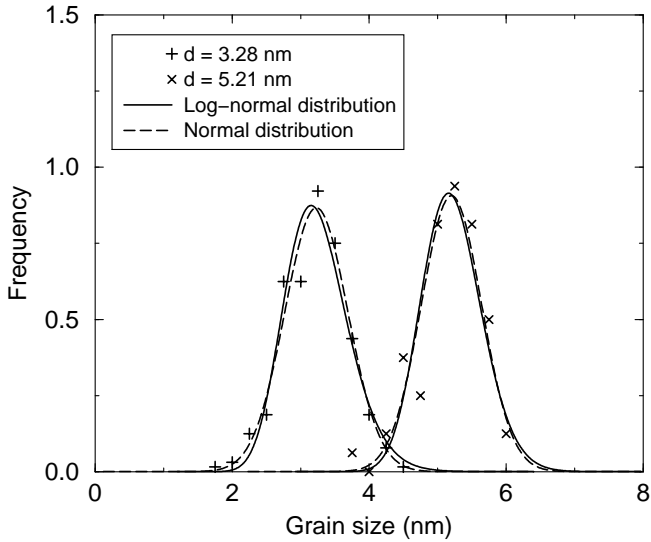, angle=-90, clip=, width=\figwidth}
  \end{center}
  \caption{The grain size distributions obtained from the Voronoi
    construction.  Grain size distributions are shown for simulations
    with intended average grain sizes of 3.28 and 5.21\,nm.  In both
    cases the input configurations from four simulations were used to
    calculate the distributions.  Fits to log-normal (solid) and normal
    (dashed) distributions are shown.  Both distributions fit the data
    well.}
  \label{fig:lognormal}
\end{figure}

\begin{figure}[htbp]
  \begin{center}
    \leavevmode
    \epsfig{file=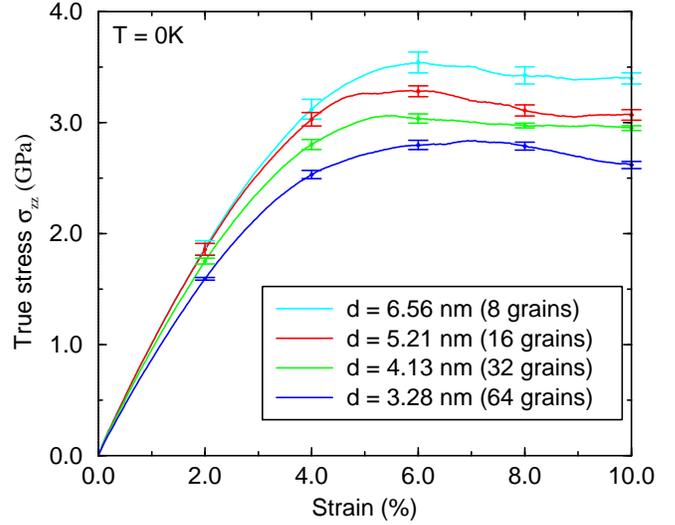, angle=-90, clip=, width=\figwidth}\\
  \end{center}
  \caption{Stress-strain curves for nanocrystalline copper at 0\,K.
    Each curve is the average over seven simulations with a given
    average grain size, the error bars indicate the uncertainty of the
    average (1$\sigma$).  Adapted from
    Ref.~\protect\onlinecite{ScDiJa98}.}
  \label{fig:stressstrain0K}
\end{figure}

\begin{figure}[htbp]
  \begin{center}
    \leavevmode
    \epsfig{file=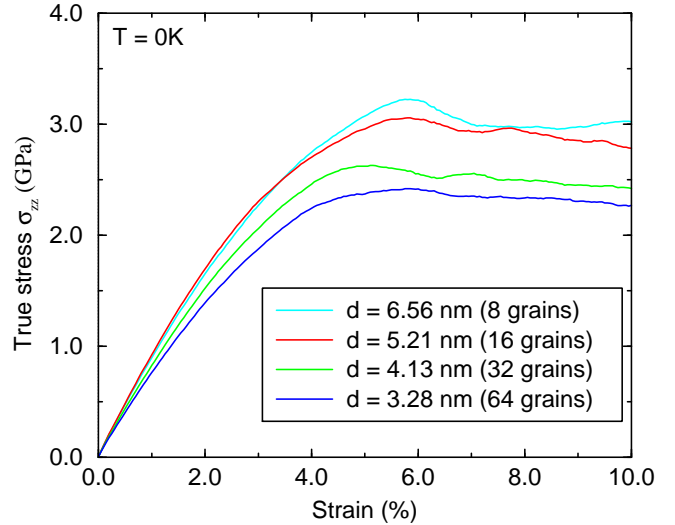, angle=-90, clip=, width=\figwidth}\\
  \end{center}
  \caption{Stress-strain curves for nanocrystalline palladium at 0\,K.
    Each curve is the average over two simulations with a given
    average grain size.}
  \label{fig:stressstrainPd}
\end{figure}

\begin{figure}[htbp]
  \begin{center}
    \epsfig{file=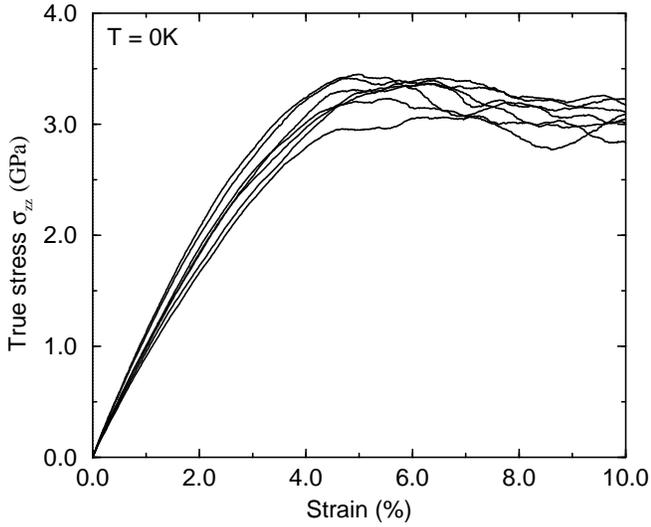, angle=-90, clip=, width=\figwidth}
  \end{center}
  \caption{Individual stress-strain curves for seven simulations of
    nanocrystalline copper with an average grain size of 5.21 nm.}
  \label{fig:stressstrainIndiv}
\end{figure}

\begin{figure}[htbp]
  \begin{center}
    \leavevmode
    \epsfig{file=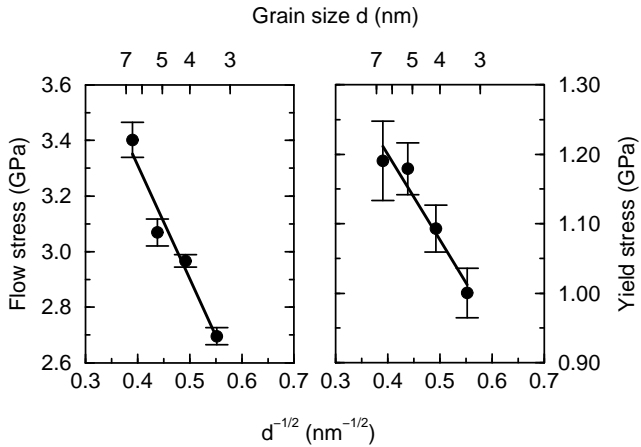, angle=-90, clip=, width=\figwidth}
  \end{center}
  \caption{Hall-Petch plot of the simulations at 0 K, showing a
    reverse Hall-Petch relationship between the grain size and both
    the yield- and flow stress.}
  \label{fig:hallpetch0K}
\end{figure}

\begin{figure}
  \begin{center}
    \leavevmode
    \epsfig{file=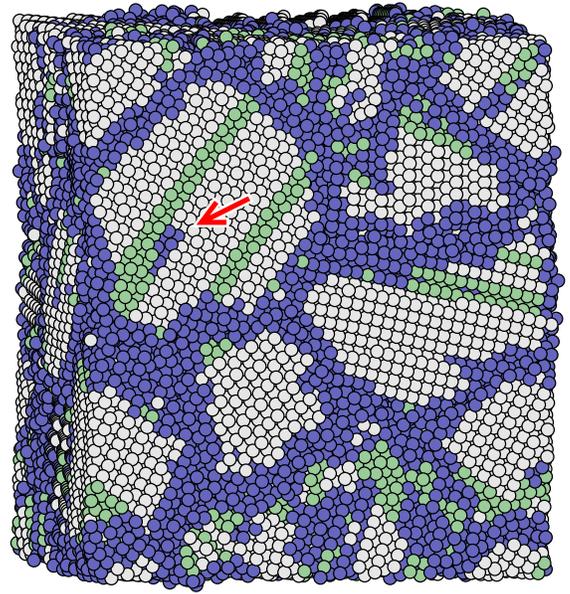, clip=, width=\figwidth}
  \end{center}
  \caption{The final configuration of the nanocrystalline copper
    system shown in Fig.~\ref{fig:initial} after 10\% deformation.
    Some stacking faults have appeared where partial dislocations have
    moved through the grains.  The arrow marks a partial dislocation
    on its way through the grain}
  \label{fig:last}
\end{figure}

\begin{figure}
  \begin{center}
    \leavevmode
    \epsfig{file=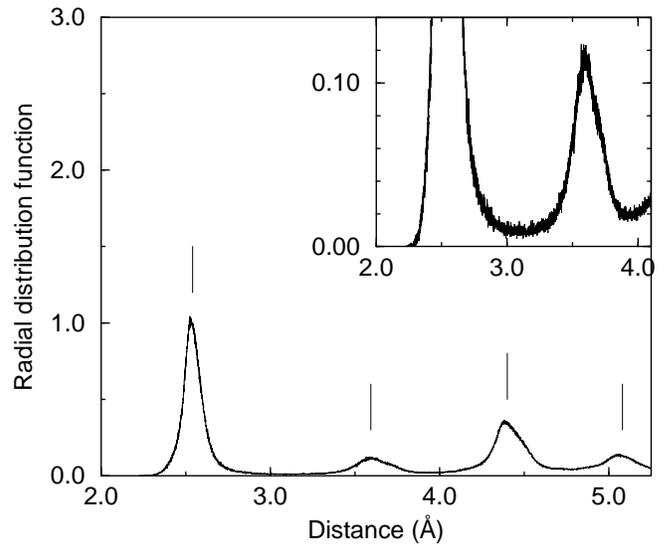, angle=-90, clip=, width=\figwidth}
  \end{center}
  \caption{The radial distribution function for the system in
    Fig.~\ref{fig:last} at $T=0$.  The marks above the
    peaks indicate the peak positions in a perfect fcc crystal.}
  \label{fig:lastrdbf}
\end{figure}

\clearpage

\begin{figure}
  \begin{center}
    \epsfig{file=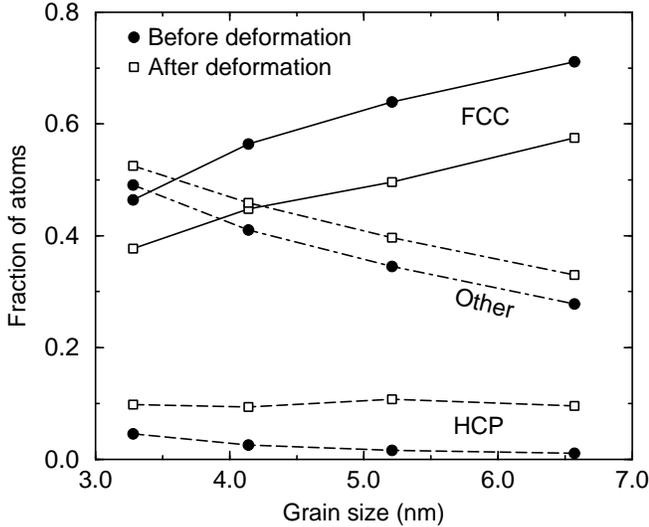, angle=-90, clip=, width=\figwidth}
    \caption{The fraction of atoms in a given local crystal structure
      as a function of the grain size.  Two curves are given for each
      type of atoms, one for the initial configurations (filled
      circles) and another showing how the fractions have changed
      after 10\% deformation (open squares).  We see how the fraction
      of atoms in the grain boundaries (marked ``Other'') decreases
      with increasing grain size, and how more atoms are in grain
      boundaries and stacking faults (hcp) after the deformation than
      before.}
    \label{fig:countatoms}
  \end{center}
\end{figure}

\begin{figure}[htbp]
  \begin{center}
    \begin{minipage}{\figwidth}
      \epsfig{file=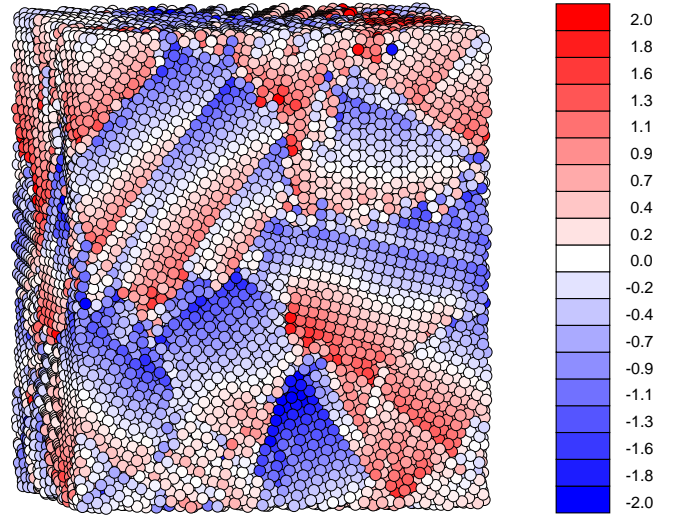, clip=, width=\figwidth}
    \end{minipage}
  \end{center}
  \caption{The same system as in Fig.~\ref{fig:last}, but with the
    atoms colors according to their motion {\em relative to the
      homogeneous deformation}.  Blue atoms have moved downwards
    relative to what would be expected if the deformation had been
    homogeneous and elastic; red atoms have moved upwards.  The scale
    indicate how far the atoms have moved (in {\AA}ngstr\"om) during
    the 10\% deformation.  The grains are seen to be blue in the upper
    part, and red in the lower part, indicating that the grains have
    not been strained as much as the system.  This indicates, that a
    major part of the deformation has been in the grain boundaries.
    Several dislocations have moved through the large grain in the
    upper-left corner.  Their slip planes are clearly seen.}
  \label{fig:motion}
\end{figure}

\begin{figure}[htbp]
  \begin{center}
    \leavevmode
    \epsfig{file=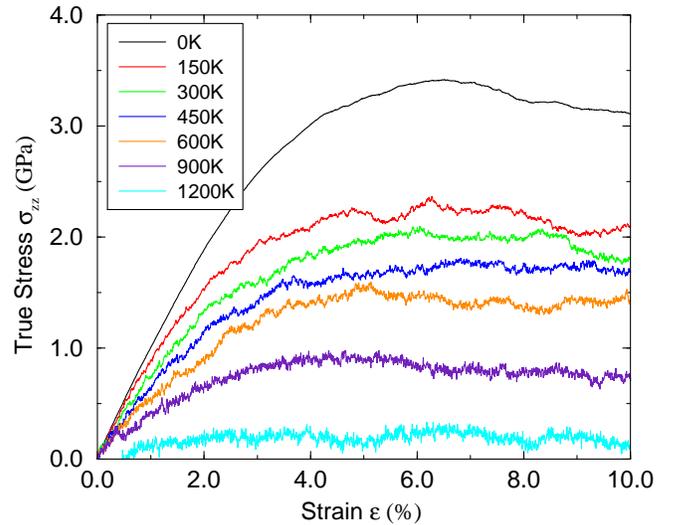, angle=-90, clip=, width=\figwidth}
  \end{center}
  \caption{Stress-strain curves for the system in
    Fig.~\protect\ref{fig:initial}, deformed at
    different temperatures.  The strain rate ($\dot\varepsilon$) is $5
    \times 10^8 s^{-1}$ except at 0\,K, where no strain rate can be
    defined (see text).  We see a clear softening at increasing temperatures.}
  \label{fig:sstemp}
\end{figure}

\begin{figure}[htbp]
  \begin{center}
    \leavevmode
    \epsfig{file=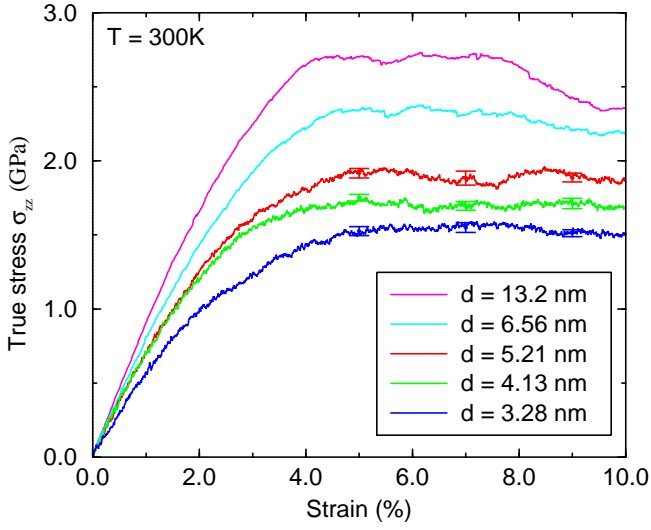, angle=-90, clip=, width=\figwidth}
  \end{center}
  \caption{Stress-strain curves for nanocrystalline copper
    at 300\,K for varying grain size.  The curves for grain sizes $d
    \le 5.21$\,nm are the average of four simulations with a system
    size of 100\,000 atoms and a strain rate of $5 \times 10^8
    s^{-1}$.  The simulations with $d \ge 6.56$\,nm were made with a
    system size of 1\,000\,000 atoms and a strain rate of $10^9
    s^{-1}$.  The influence of this change in the strain rate is
    minimal, see section \protect\ref{sec:strainrate} and
    Fig.~\protect\ref{fig:strainrate}.  One simulation was performed
    with $d = 6.56$\,nm, two with $d = 13.2$\,nm.  The thermal
    fluctuations are less pronounced in the simulations with the
    larger system size.}
  \label{fig:stressstrain300K}
\end{figure}

\begin{figure}[htbp]
  \begin{center}
    \leavevmode
    \epsfig{file=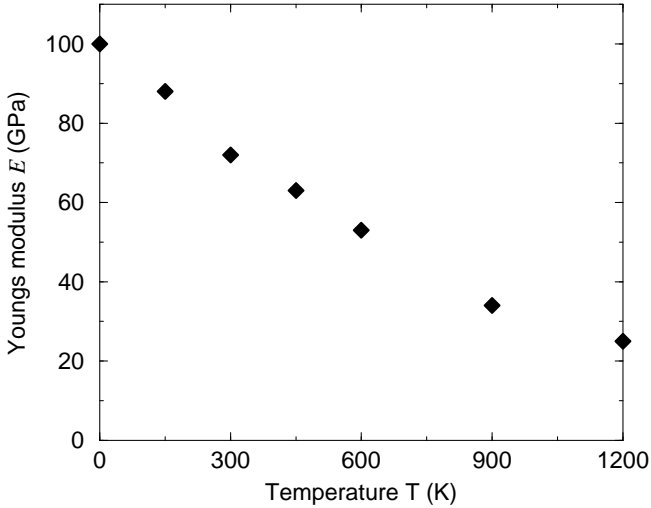, angle=-90, clip=, width=\figwidth}
  \end{center}
  \caption{Young's modulus as a function of temperature for a sample
    with an average grain size of 5.2 nm.}
  \label{fig:youngtemp}
\end{figure}

\begin{figure}[htbp]
  \begin{center}
    \leavevmode
    \epsfig{file=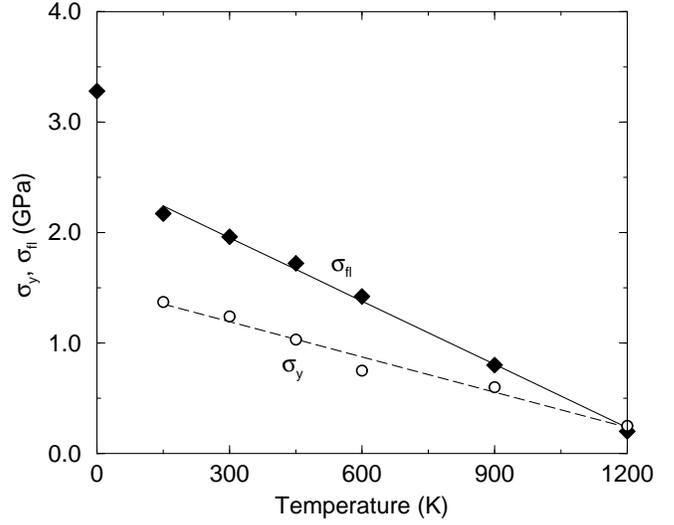, angle=-90, clip=, width=\figwidth}
  \end{center}
  \caption{Yield and flow stresses as a function of temperature.  No
    yield stress is given for $T=0 K$, as the yield stress at this
    temperature cannot directly be compared to that at finite
    temperatures, see section \ref{sec:youngfinite}.}
  \label{fig:flowtemp}
\end{figure}

\begin{figure}[htbp]
  \begin{center}
    \epsfig{file=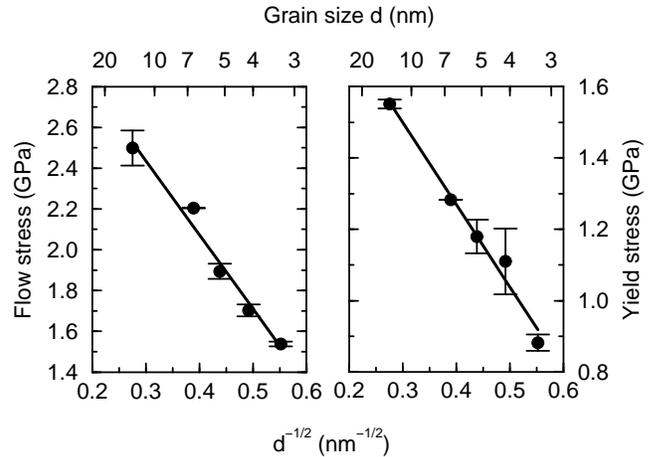, angle=-90, clip=, width=\figwidth}
  \end{center}
  \caption{A Hall-Petch plot of the simulations at 300K.  As at 0K, a reverse
    Hall-Petch relation is seen.}
  \label{fig:hallpetch300K}
\end{figure}

\begin{figure}[htbp]
  \begin{center}
    \begin{minipage}{\figwidth}
      \epsfig{file=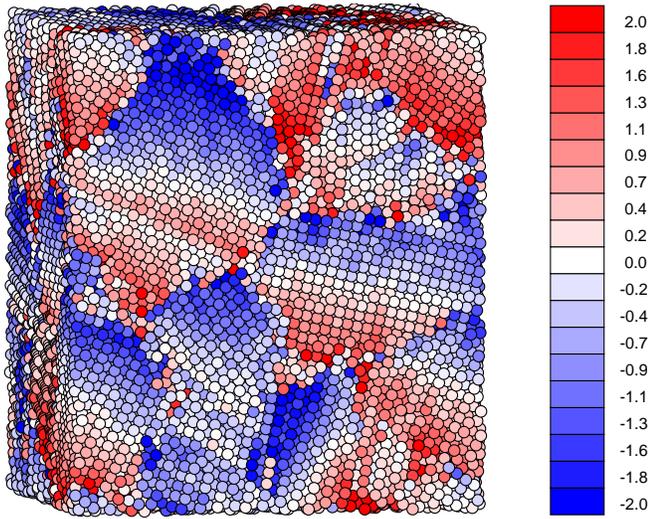, clip=, width=\figwidth}
    \end{minipage}
  \end{center}
  \caption{The same system as in Fig.~\ref{fig:motion}, but after
    deformation at 300\,K, see the caption of that figure for a
    description of the color coding.  The main deformation is
    seen to be in the grain boundaries, as was the case  at 0\,K.}
  \label{fig:motion300K}
\end{figure}

\begin{figure}[htbp]
  \begin{center}
    \leavevmode
    \epsfig{file=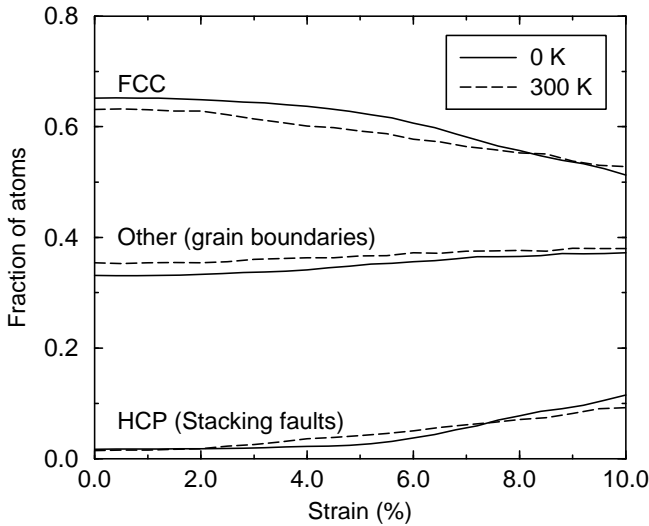, angle=-90, clip=, width=\figwidth}
  \end{center}
  \caption{Development of the number of atoms in different local
    environments as a function of the strain, for a single simulation
    at 0\,K and at 300\,K.}
  \label{fig:classdevel}
\end{figure}

\begin{figure}[htbp]
  \begin{center}
    \leavevmode
    \epsfig{file=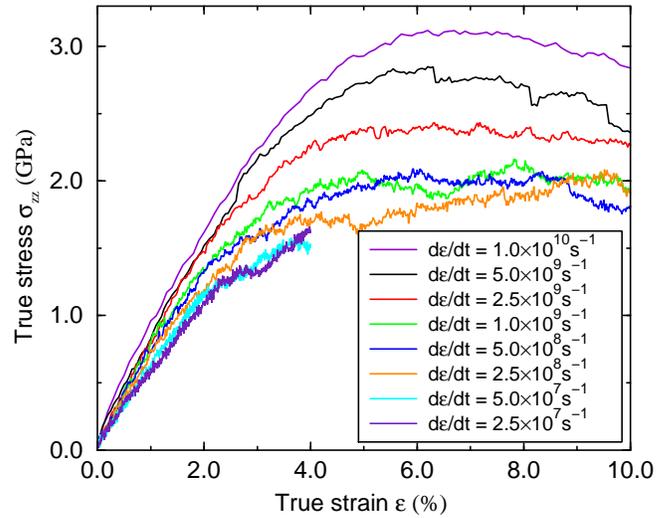, angle=-90, clip=, width=\figwidth}
  \end{center}
  \caption{The effect of varying the strain rate.  Stress-strain
    curves are shown for simulations of the same system at the same
    temperature (300 K), but with varying strain rates.  A strong
    influence of strain rate is seen for strain rates above
    $\dot\varepsilon = 10^9 s^{-1}$, below that value the strain rate
    dependence is less pronounced.  The simulations at the two lowest
    strain rates were stopped after 4\% deformation.}
  \label{fig:strainrate}
\end{figure}

\begin{figure}[htbp]
  \begin{center}
    \leavevmode
    \epsfig{file=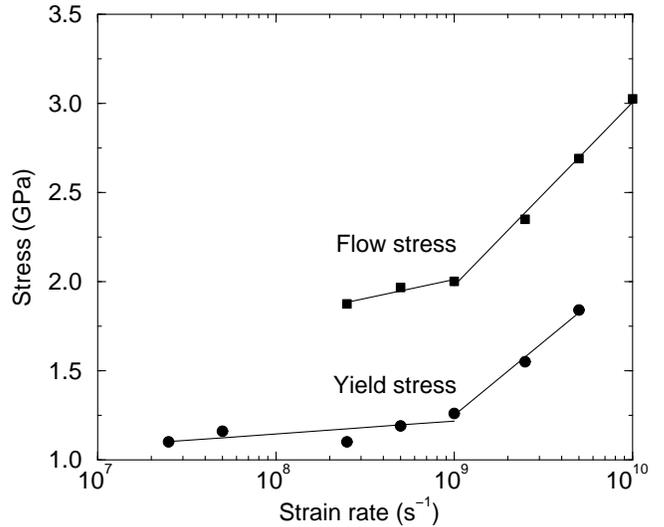, angle=-90, clip=, width=\figwidth}
  \end{center}
  \caption{Summary of the effect of varying the strain rate.  Both the
    yield stress and the flow stress are seen to vary with the strain
    rate ($\dot\varepsilon$), strongest for $\dot\varepsilon > 10^9
    s^{-1}$.}
  \label{fig:sratesummary}
\end{figure}

\begin{figure}[htbp]
  \begin{center}
    \leavevmode
    \epsfig{file=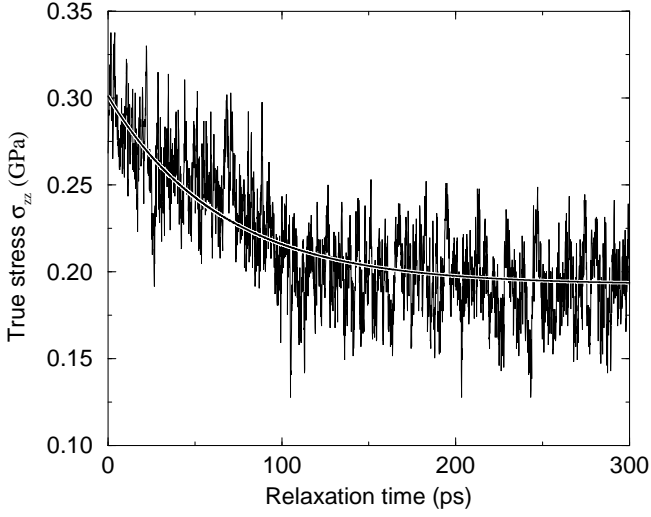, angle=-90, clip=, width=\figwidth}
  \end{center}
  \caption{Sample relaxation in the ``elastic'' region.  The
    deformation is stopped after 0.4\% deformation at $\dot\varepsilon
    = 2.5\times10^7 s^{-1}$ and $T = 300 K$.  The stress is seen to
    decrease with a characteristic time of $\sim$100 ps, stabilizing
    on a value of $2/3$ the original stress.  The thick black line is an
    exponential fit.}
  \label{fig:relaxation}
\end{figure}

\begin{figure}[htbp]
  \begin{center}
    \epsfig{file=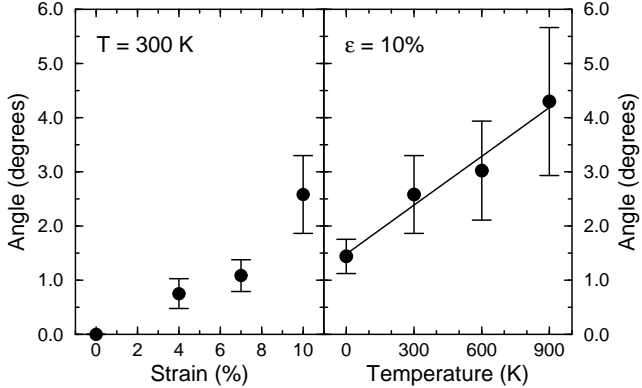, angle=-90, clip=, width=\figwidth}
    \caption{Grain rotation as a function of strain and temperature.
      The rotation of five randomly selected grains were followed in
      simulations at different temperatures.  The average grain size
      is 5.2 nm.  The error bars indicate the $1\sigma$ spread in the
      rotations.}
    \label{fig:rotation}
  \end{center}
\end{figure}

\begin{figure}[htbp]
  \begin{center}
    \leavevmode
    \epsfig{file=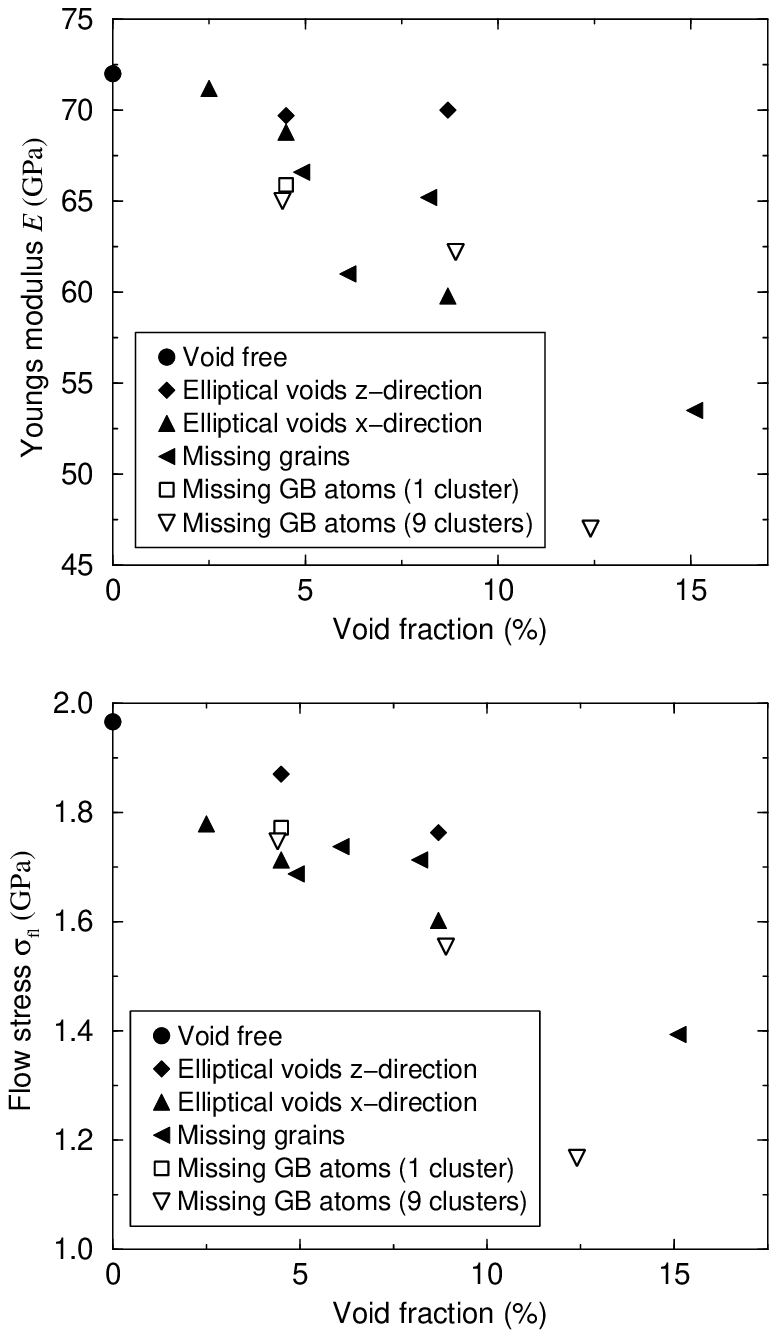, clip=, width=\figwidth}
  \end{center}
  \caption{Effect of voids on the Young's Modulus (top) and on the
    flow stress (bottom), showing a decrease of both quantities with
    increasing void fraction.  The voids were generated by removing
    selected atoms, the ``void fraction'' is the fraction of atoms
    removed.  Different methods were used to select the atoms to be
    removed, see section \ref{sec:porosity}.}
  \label{fig:voids}
\end{figure}

\end{document}